\documentclass[aps,pra,twocolumn,superscriptaddress]{revtex4}
\usepackage[latin1]{inputenc}
\usepackage[T1]{fontenc}
\usepackage{amsmath}
\usepackage{hyperref}
\usepackage{sistyle}
\usepackage{graphicx}
\graphicspath{{./Figures/}}
\usepackage{subfigure}
\usepackage{nicefrac}
\usepackage{framed,color}
\definecolor{shadecolor}{gray}{0.85}

\newcommand{\paren}[1]{\left( #1 \right)}

\newcommand{\brc}[1]{\left\{ #1 \right\}}

\newcommand{\abs}[1]{\left| #1 \right|}

\newcommand{\ket}[1]{\left| #1 \right\rangle}

\usepackage{array}

\newcommand{\perc}[1]{\SI{#1}{\%}}

\begin{document}
\title{Resilience of orbital angular momentum qubits and effects on hybrid entanglement}
\author{Daniele Giovannini}
\affiliation{Dipartimento di Fisica, Sapienza Universit\`a di Roma, Roma 00185, Italy}
\author{Eleonora Nagali}
\affiliation{Dipartimento di Fisica, Sapienza Universit\`a di Roma, Roma 00185, Italy}
\author{Lorenzo Marrucci}
\affiliation{Dipartimento di Scienze Fisiche, Universit\`a di Napoli ``Federico II'', Complesso Universitario di Monte S. Angelo, 80126 Napoli, Italy}
\affiliation{CNR-SPIN, Complesso Universitario di Monte S.~Angelo, 80126 Napoli, Italy}
\author{Fabio Sciarrino}
\email{fabio.sciarrino@uniroma1.it}
\affiliation{Dipartimento di Fisica, Sapienza Universit\`a di Roma, Roma 00185, Italy}
\affiliation{Istituto Nazionale di Ottica (INO-CNR), L.go E.~Fermi 6, Florence 50125, Italy}

\begin{abstract}
The orbital angular momentum of light (OAM) provides a promising
approach for the implementation of multidimensional states (qudits)
for quantum information purposes. In order to characterize the
degradation undergone by the information content of qubits encoded
in a bidimensional subspace of the orbital angular momentum degree
of freedom of photons, we study how the state fidelity is affected
by a transverse obstruction placed along the propagation direction
of the light beam. Emphasis is placed on the effects of planar and
radial hard-edged aperture functions on the state fidelity of
Laguerre-Gaussian transverse modes and the entanglement properties
of polarization-OAM hybrid-entangled photon pairs.
\end{abstract}

\maketitle

\section{Introduction}
In quantum information theory the fundamental unit of information is
a two-level quantum system, the qubit. As in classical information
science with bits, all quantum information tasks can at least in
theory be performed through just qubits and quantum gates operating
on qubits \cite{Nielsen:2000, Gisin:2002, OBrien:2007}. For quantum
information purposes and the effective production and processing of
robust qubits, as well as that of multi-dimensional quantum states
or qudits \cite{Nagali:2010a, Nagali:2010prl}, considerable interest
has been recently focused on the generation and manipulation of
helical laser beams. These optical waves, which have been shown to
carry well-defined values of orbital angular momentum (OAM), are
well described in terms of Laguerre-Gaussian (LG) modes, containing
an \(\ell\)-charged optical phase singularity (or optical vortex) at
their beam axis \cite{Mair:2001, Molina-Terriza:2008,
Franke-Arnold:2008}. Any two of such OAM modes with opposite value
of $\ell$ and a common radial profile, here denoted as $\ket{+\ell}$
and $\ket{-\ell}$, define a basis of a bidimensional OAM subspace
$o_{|\ell|}$ in which one can encode a generic qubit. These specific
OAM subspaces are particularly convenient to this purpose as they
are not affected by propagation-induced decoherence, because the
radial profile factorizes and can be usually ignored
\cite{Naga:2009opt}. Experimentally, single-photon qubits in the OAM
subspace \(o_2\) can be efficiently encoded and read-out by means of
``polarization-OAM transferrers'' \cite{Naga:2009prl, Nagali:2009d},
which are devices based on a recently introduced optical element
called ``q-plate'' \cite{Marrucci:2006prl}.

The growing number of works demonstrating simple quantum information
protocols based on OAM-encoded qubits raise questions about how
practical is such approach, as compared for instance with the
standard polarization encoding. For example, one can ask how
sensitive are these OAM-encoded qubits to small optical
misalignments or other non-ideality of typical setups. More in
general, studies of the resilience of OAM-encoded qubits or qudits
in free-space propagation under the effect of perturbations and of
imperfect detection are pivotal for the use of the OAM of light for
quantum communication tasks. Previous works on LG modes in the
classical regime showed for example the spread of measured
transverse modes when planar restrictions spanning an angular range
of less than \(2\pi\) are placed along the beam
\cite{Arlt:2003,Gibson:2004}, the consequences of non-coaxial mode
detection \cite{Vasnetsov:2005}, and the effect of turbulence on
OAM-encoded information
\cite{Paterson:2005,Gopaul:2007,Gbur:2008,Pors:2009,Tyler:2009,Semenov:2010}.

\begin{figure}
\subfigure[\(B(x_0,x;y)\)]{\includegraphics[width=0.43\linewidth]{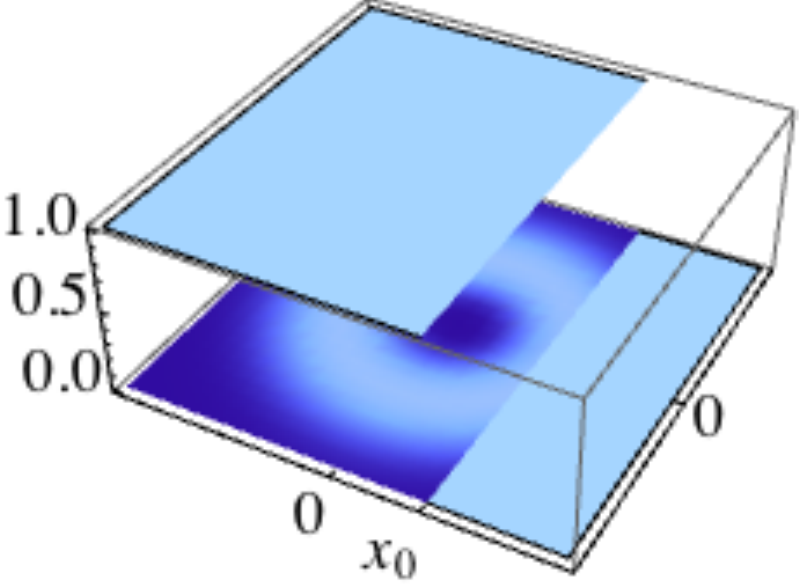}\label{fig:ApertureX}} \quad%
\subfigure[\(\Pi(r_0,r;\phi)\)]{\includegraphics[width=0.43\linewidth]{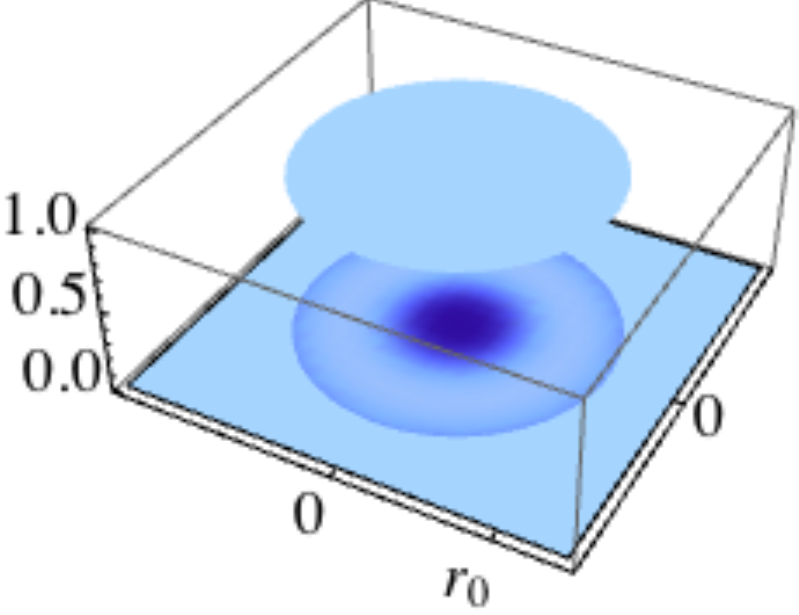}\label{fig:ApertureR}}
\caption{\label{fig:Aperture} Aperture functions:
\subref{fig:ApertureX} planar transverse obstruction \(B(x_0,x;y)\)
(knife), and \subref{fig:ApertureR} radial obstruction
\(\Pi(r_0,r;\phi)\) (iris).}
\end{figure}

The purpose of this paper is to characterize the degradation
undergone by the information encoded in \(o_2\) OAM qubits when
transverse hard-edged optical apertures are placed along the
propagation direction of the beam, both in the classical and quantum
regimes. The apertures considered in this work have one of the two
shapes shown in Fig.~\ref{fig:Aperture}, which are here taken to
mimic the typical effect of optical misalignments or of the finite
numerical aperture of optical components.

This paper is organized as follows: Sec.~\ref{sec:Theory} introduces
a classical theoretical model, based on the functional shape of LG
modes, used to assess the degradation undergone by states belonging
to three different \(o_2\) bases as a consequence of the
perturbation introduced during propagation by the two kinds of
optical apertures. In Sec.~\ref{sec:ClassicalExperimental}, we
compare the theoretical predictions with the results of experiments
performed in the classical regime. In
Sec.~\ref{sec:HybridEntanglement} we finally present the
experimental results concerning the effects of the previously
introduced obstructions on the entanglement properties of a hybrid
polarization-OAM entangled pair of twin photons \cite{Nagali:2010,
Karimi:2010glasgow}, that is, a pair of entangled photons whose entanglement is encoded in two different degrees of freedom.

\section{Theory}
\label{sec:Theory} An OAM eigenstate $\ket{\ell}$ is here taken to
denote a pure LG mode, corresponding to a wavefunction
\(A(x,y,z)=A(r,\phi,z)=A_0(r,z) \, e^{i\ell\phi}\) with azimuthal
index $\ell$, where $z$ is the propagation axis, $x,y$ are the
cartesian coordinates for the transverse plane, and $r,\phi$ the
corresponding polar coordinates. LG modes are characterized also by
another index $p$, determining the radial profile. In the following,
except where explicitly stated otherwise, this radial index is
understood to be $p=0$ and will be omitted.

Generic OAM qubits are described by superpositions
$\ket{\psi}=\alpha\ket{+2}+\beta\ket{-2}$, where $\alpha$ and
$\beta$ are complex coefficients. We will consider in particular the
following six representative states, belonging to three mutually
unbiased bases of the OAM subspace $o_2$: \(\ket{l}=\ket{+2}\),
\(\ket{r}=\ket{-2}\); \(\ket{h}=\paren{\ket{l}+\ket{r}}/\sqrt{2}\),
\(\ket{v}=-i\paren{\ket{l}-\ket{r}}/\sqrt{2}\); and
\(\ket{d}=(1-i)\paren{\ket{l}+i\ket{r}}/2\),
\(\ket{a}=(1+i)\paren{\ket{l}-i\ket{r}}/2\) \cite{Padgett:1999}. The
wavefunctions $A(x,y,z)$ of these states are given by the
corresponding superpositions of pure LG modes. All input
wavefunctions are normalized for integration in any given
(arbitrary) transverse plane $z$ of the beam, i.e., $\iint\abs{A}^2
dx\,dy =1$.

As mentioned in the Introduction, we consider two sets of optical
apertures: (i) a half-plane obstruction with its edge located at the
variable abscissa $x_0$, described by the transmittance function
\(B(x_0,x;y) = \theta(x_0-x)\) where \(\theta(x)\) is the Heaviside
step function (see Fig.~\ref{fig:ApertureX}); and (ii) an iris with
variable aperture radius \(r_0\), described by the transmittance
function \(\Pi(r_0,r;\phi) = \theta(r_0-r)\) (see
Fig.~\ref{fig:ApertureR}). The first kind of aperture is meant to
mimic the perturbations arising from small transverse optical
misalignments, while the second corresponds to introducing optical
elements having a small numerical aperture.
\begin{figure}
\includegraphics[width=0.6\linewidth]{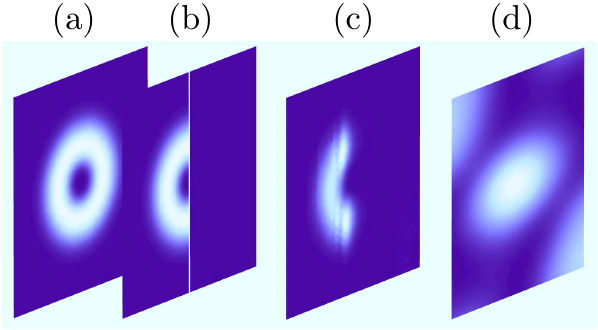}
\caption{Propagation of an \(\text{LG}_{0,+2}\) beam (basis
state~\(\ket{l}\)) through a \(B(x_0)\) aperture function with
\(T=\num{0.5}\). Profiles are
drawn to the same scale. From left to right: (a) unperturbed transverse
intensity profile; (b) intensity profile immediately after the
aperture function (0.01 nm from the obstruction); (c) calculated profile after free-space
propagation of the perturbed mode for $d=30$cm after the obstruction; (d) calculated profile after the
insertion of a q-plate optical element (see Ref.\
\cite{Marrucci:2006prl}), used for the conversion into a
\(\text{TEM}_{00}\) mode after the propagation of $d=30$cm.} \label{fig:BProp}
\end{figure}

\begin{figure}
\subfigure[\(T=\num{1}\)]{\includegraphics[width=0.49\linewidth]{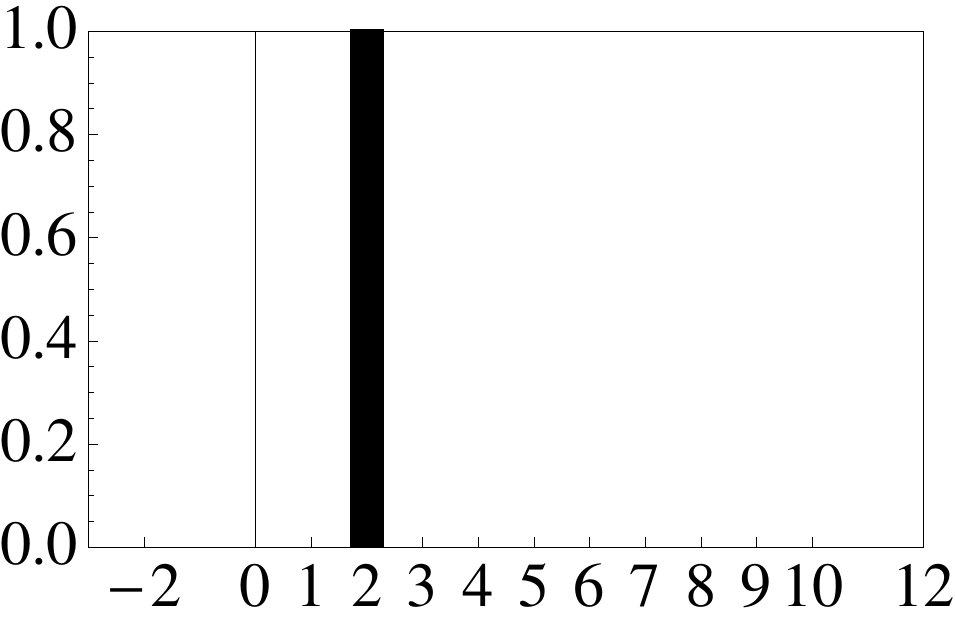}} \hfill%
\subfigure[\(T=\num{0.5}\)]{\includegraphics[width=0.49\linewidth]{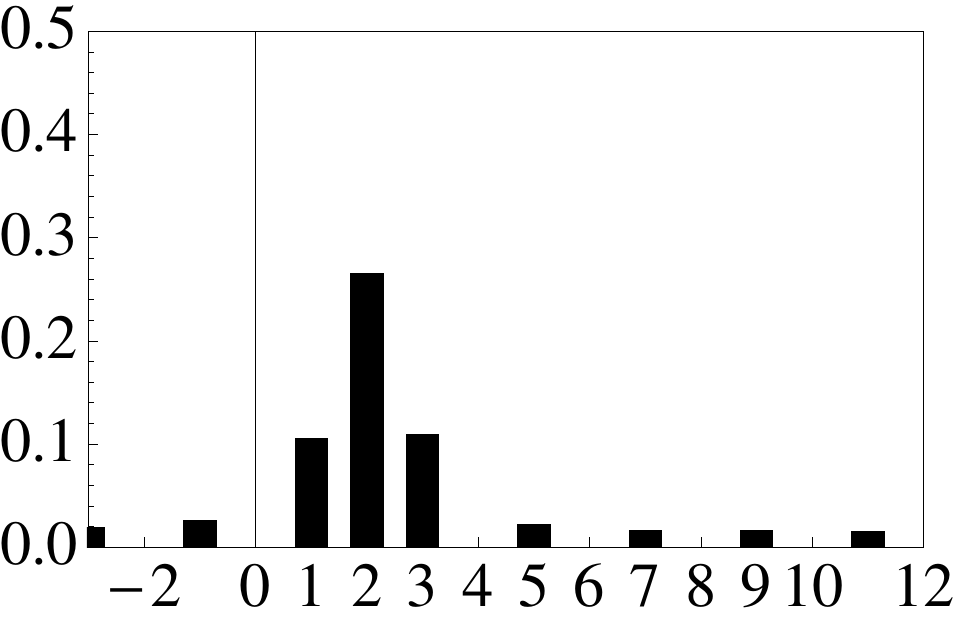}} \\
\subfigure[\(T=\num{0.28}\)]{\includegraphics[width=0.49\linewidth]{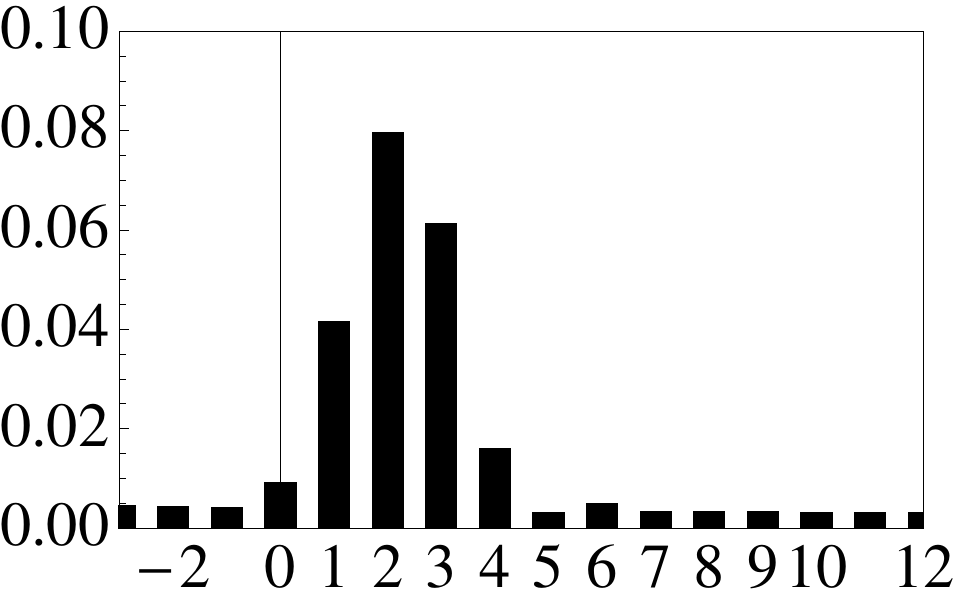}} \hfill%
\subfigure[\(T=\num{0.05}\)]{\includegraphics[width=0.49\linewidth]{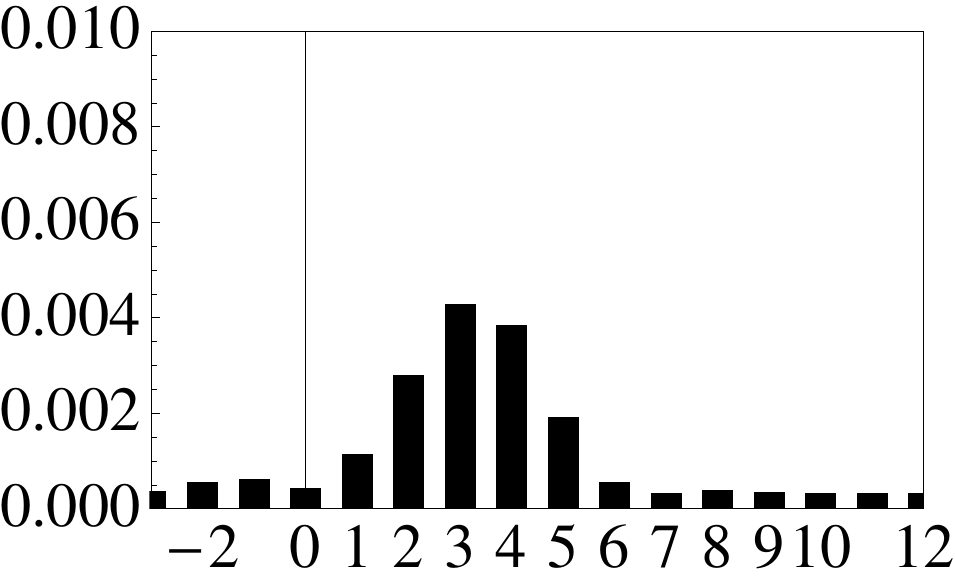}}
\caption{Spread in the measurement probabilities of OAM modes with
\(\ell^\prime=-2,\dots 12\) for various positions \(x_0\) of a
\(B(x_0)\) aperture inserted into the path of an \(\ell=2\) beam
(i.e. for decreasing values of transmittance \(T\)).}
\label{fig:Spectrum}
\end{figure}

Now, given an arbitrary input qubit $\ket{\psi}$ as described by the
wavefunction $A$, the perturbed state $\ket{\psi^\prime}$ obtained
immediately after the aperture is given by a (not normalized)
wavefunction \(A^\prime=A\cdot B\) or \(A^\prime=A\cdot \Pi\),
respectively (see, e.g., Fig.\ \ref{fig:BProp}a-b). Given the
normalization of the input wavefunction, the transmitted fraction of
photons after the aperture is given by \(T=\iint\abs{A^\prime}^2
dx\,dy\). We then introduce the following two projections:
\begin{subequations}
\begin{align}
\kappa_\psi &= \displaystyle \int_{-\infty}^{\infty} \!\!\!\!\!\! dx \!\! \int_{-\infty}^\infty \!\!\!\!\!\! dy \; A^\ast(x,y) \; A^\prime(x,y) \\
\kappa_{\psi^\perp} &= \displaystyle \int_{-\infty}^{\infty}
\!\!\!\!\!\! dx \!\! \int_{-\infty}^\infty \!\!\!\!\!\! dy
\paren{A^\perp(x,y)}^\ast A^\prime(x,y),
\end{align}
\end{subequations}
where \(A^\perp\) is the wavefunction of the orthogonal state within
the $o_2$ OAM subspace (with $p=0$). We take the final detection
probability of states \(\psi\) and \(\psi^\perp\) to be given by
\(P(\psi) = |\kappa_\psi|^2\) and
\(P(\psi^\perp)=|\kappa_{\psi^\perp}|^2\), respectively. Therefore,
\(P_{o_2}=|\kappa_\psi|^2+|\kappa_{\psi^\perp}|^2\) represents the
probability of information preservation in the OAM subspace \(o_2\)
(with $p=0$) after the initial mode has been transmitted through the
aperture. The complementary fraction $1-P_{o_2}$ of photons is lost
either because they are absorbed in the aperture (as given by the
fraction $1-T$) or because they are transferred out of the $o_2$
$p=0$ subspace, and therefore are finally filtered out by the
measurement process (note that this implies that the detection
optics is assumed to have a small numerical aperture, thus strongly
favoring $p=0$ over higher radial modes). The computed OAM spectrum
broadening induced by the planar aperture (function $B$) is shown in
Fig.\ \ref{fig:Spectrum}, for different positions of the aperture as
specified by the resulting transmittance $T$.

The detected photons, however, will transport a partially degraded
OAM qubit. The amount of quantum information content that can be
reconstructed by a receiver by means of projective measurements can
be described through a fidelity parameter \(F\) defined as
\(F=P(\psi)/[P(\psi)+P(\psi^\perp)]\).

\begin{figure}
\subfigure[\(\ket{l}\)]{\includegraphics[width=0.49\linewidth]{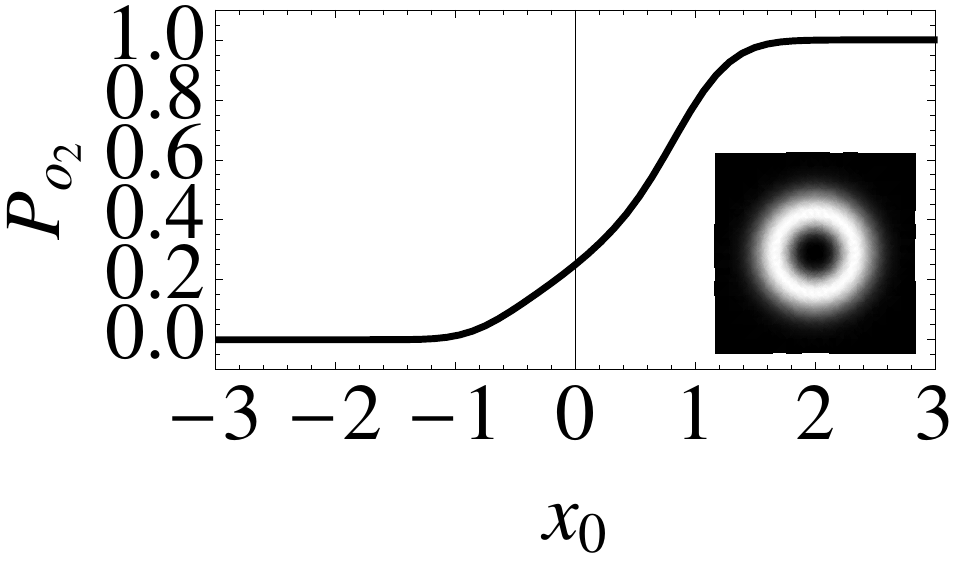}}
\subfigure[\(\ket{h}\)]{\includegraphics[width=0.49\linewidth]{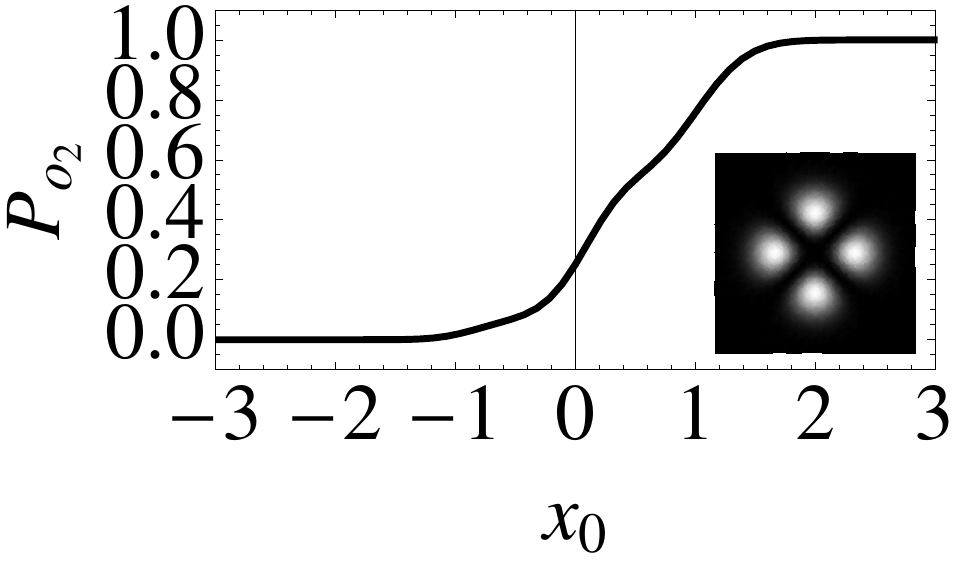}} \\
\subfigure[\(\ket{v}\)]{\includegraphics[width=0.49\linewidth]{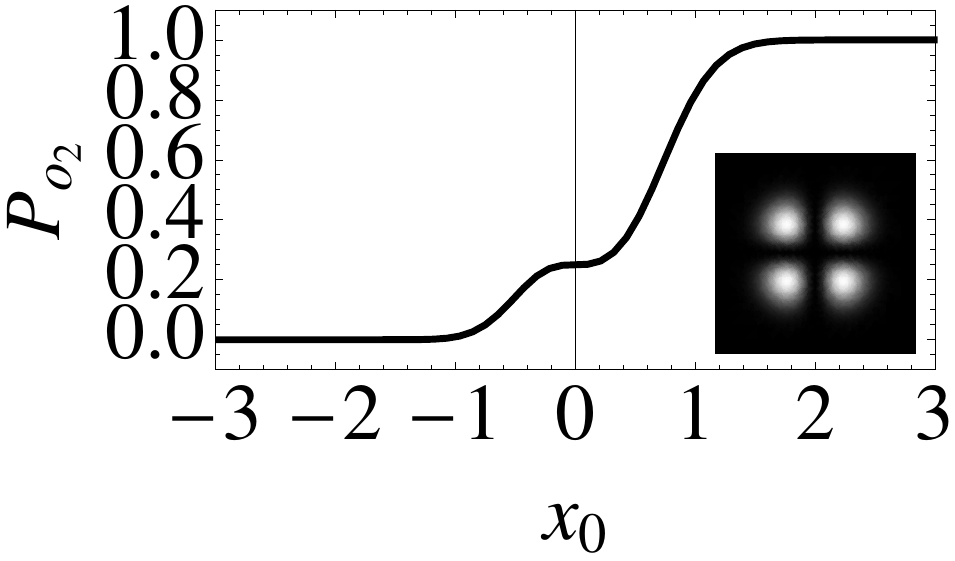}}
\subfigure[\(\ket{d}\)]{\includegraphics[width=0.49\linewidth]{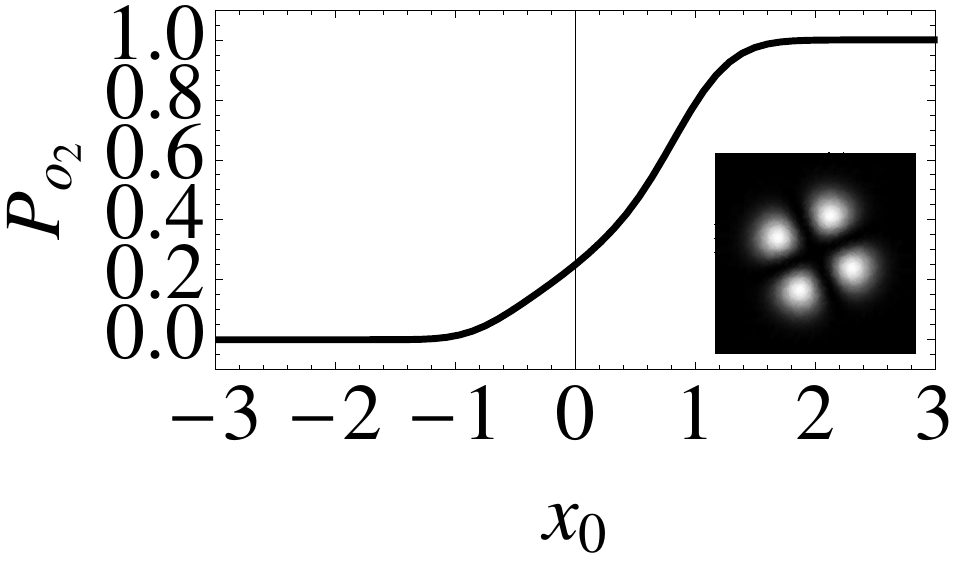}}
\caption{Theoretical curves of \(P_{o_2}\) for aperture function
\(B(x_0)\) for input states \(\ket{l}\) (\(\ket{r}\) exhibits the
same curve), \(\ket{h}\), \(\ket{v}\) and \(\ket{d}\) (or
\(\ket{a}\)). The aperture edge position \(x_0\) is given in units
of the beam waist \(w_0\)). The mode intensity profile of each state
is also shown in the inset of each panel. The aperture edge is
vertically oriented in the inset frames.} \label{fig:PlotsThB}
\end{figure}

For each OAM state \(\ket{\psi}\), and each position \(x_0\) (or
radius \(r_0\)) of the obstruction, the probabilities \(P(\psi)\)
and \(P(\psi^\perp)\) were computed using numerical integration. The
corresponding predictions for the information preservation
probability \(P_{o_2}\) are shown in Fig.\ \ref{fig:PlotsThB},
together with the corresponding mode profiles.

The computed mean information preservation probability \(P_{o_2}\)
and fidelity \(F\) for the aperture function \(B(x_0)\), averaged
over the six OAM states, are shown in Fig.~\ref{fig:PlotsB} as solid
lines. The mean \(P_{o_2}\) relative to the aperture function
\(P(r_0)\) is shown in Fig.~\ref{fig:PlotrP}. The theoretical
fidelity in this second case is constantly unitary, because the
rotational symmetry is not broken by the aperture.

\section{Classical experiments} \label{sec:ClassicalExperimental}
Classical measurements were performed on coherent beams by
generating the OAM modes in \(o_2\) corresponding to the six states
(\(\ket{l}\), \(\ket{r}\)), (\(\ket{h}\), \(\ket{v}\), \(\ket{d}\),
\(\ket{a}\)), and then placing either a knife (aperture function
\(B\)) or an iris (function $\Pi$) along the beam path. The
perturbed modes \(\ket{\psi^\prime}\) were then projected onto
\(\ket{\psi}\) and \(\ket{\psi^\perp}\) by the analysis setup. We
assumed that no significant degradation (except possibly for some
additional small losses) occurs to the quantum information in the
analysis setup.

In the experimental implementation, a continuous laser beam was
coupled to a single-mode fiber in order to collapse its transverse
spatial mode into a pure \(\text{TEM}_{00}\), corresponding to OAM
\(\ell=0\). After the fiber, a polarization set was used to prepare
the input polarization state as one of the six qubits. The beam was
then sent through a quantum transferrer \(\pi\to o_2\), which
transferred the polarization quantum state to the OAM degree of
freedom, thus obtaining one of the input OAM state to be studied
\cite{Naga:2009opt,Naga:2009prl}. In order to analyze with high
efficiency the OAM state after the aperture, we then exploited an
inverse \(o_2\to\pi\) transferrer and a polarization analysis set
\cite{Naga:2009opt}. The inverse transferrer includes the coupling
to a single-mode fiber, which in our case was achieved with a
typical efficiency \(\eta=\perc{14.8}\). Although the q-plate device
used in the transferrers is known to generate a radial profile more
complex than a pure LG mode \cite{Karimi:2009ol}, only the $p=0$
radial modes will be efficiently coupled to the final single-mode
fiber used in our analysis setup, so that the $p=0$ assumption used
in our theory appears to be well justified.

\begin{figure}
\subfigure[][]{\includegraphics[width=\linewidth]{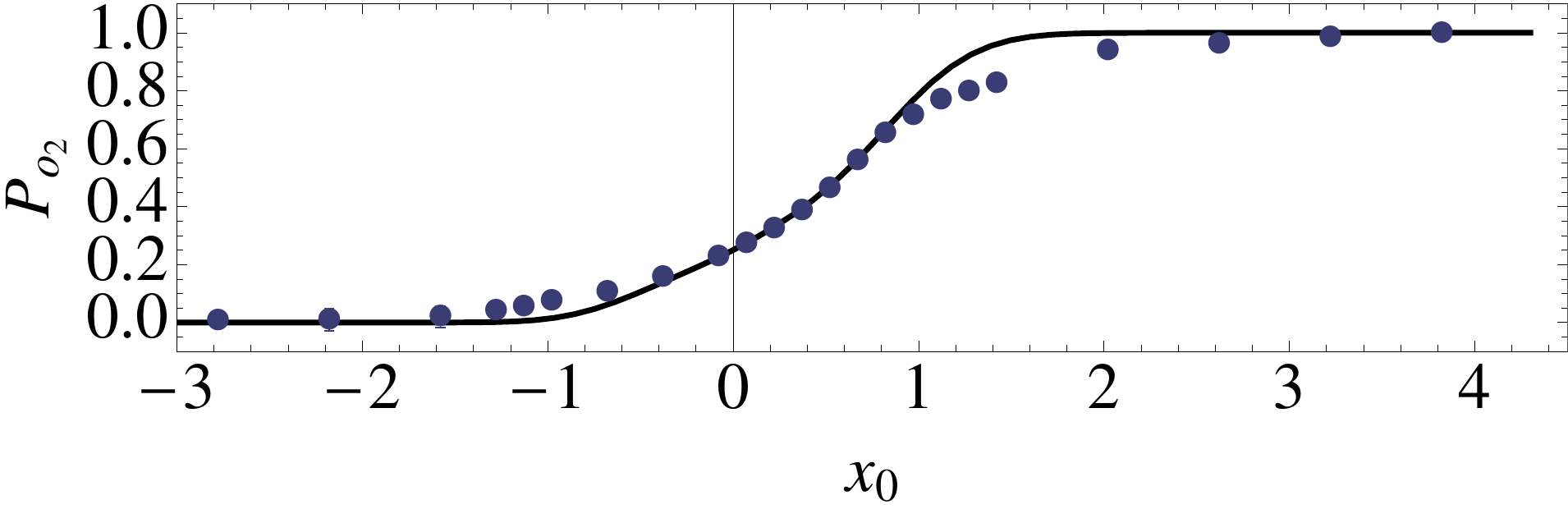}\label{fig:PlotxP}} \\
\subfigure[][]{\includegraphics[width=\linewidth]{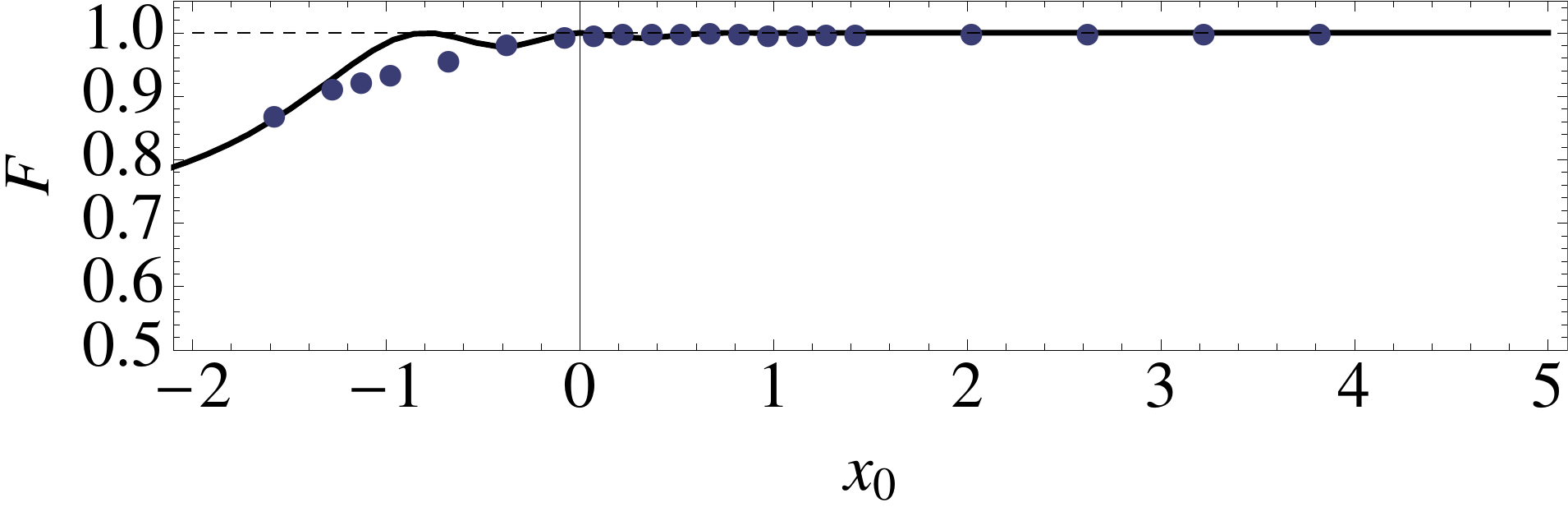}\label{fig:PlotxF}}
\caption{Average information preservation probability \(P_{o_2}\)
(\textbf{a}) and fidelity (\textbf{b}) for planar aperture function
\(B(x_0)\): lines are theoretical predictions, circles are data
obtained in classical measurements (average over states \(\ket{d}\)
and \(\ket{a}\)); experimental errors are negligible. The aperture
edge position \(x_0\) is given in units of the beam waist \(w_0\)).}
\label{fig:PlotsB}
\end{figure}

\begin{figure}
\subfigure[][]{\includegraphics[width=\linewidth]{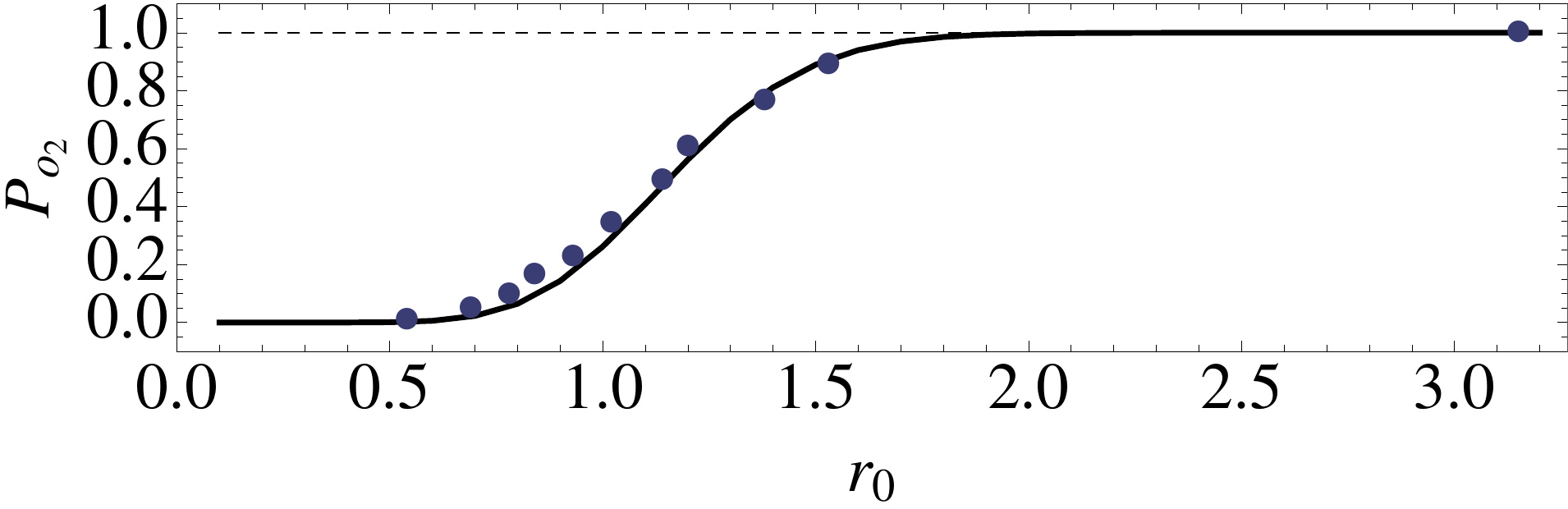}\label{fig:PlotPrP}} \\
\subfigure[][]{\includegraphics[width=\linewidth]{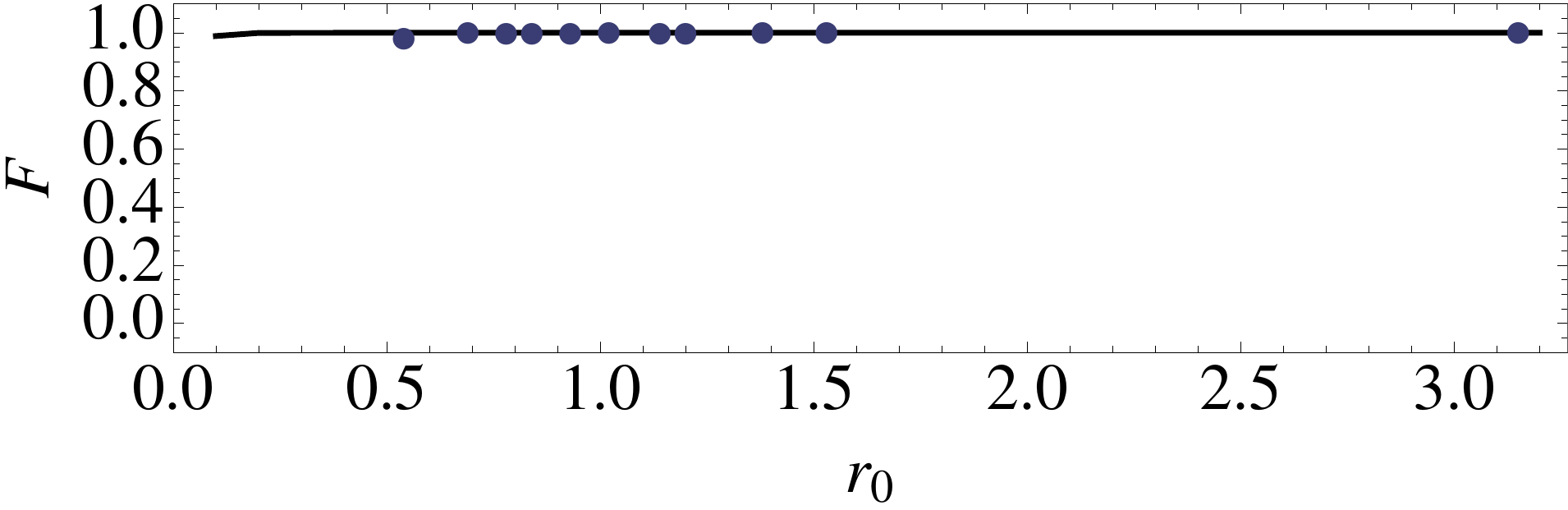}\label{fig:PlotPrF}}
\caption{Average information preservation probability \(P_{o_2}\)
(\textbf{a}) and fidelity (\textbf{b}) for circular aperture
function \(\Pi(r_0)\): lines are theoretical predictions, circles
are data obtained in classical measurements, as a function of the
aperture radius \(r_0\) in units of the beam waist \(w_0\)).}
\label{fig:PlotrP}
\end{figure}

The experimental average data for aperture \(B\) and states
\(\ket{d}\) and \(\ket{a}\) are shown in Fig.~\ref{fig:PlotsB},
those for \(\Pi\) and all \(o_2\) basis states in
Fig.~\ref{fig:PlotrP}.


\begin{figure}
\subfigure[$B(x_0)$]{\includegraphics[width=0.49\linewidth]{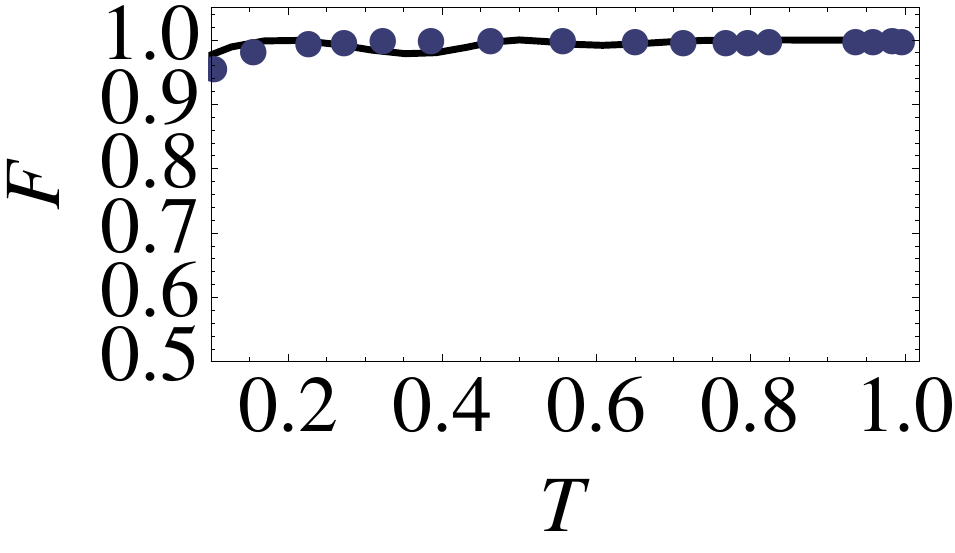}\label{fig:PlotBTF}}
\subfigure[\(\Pi(r_0)\)]{\includegraphics[width=0.49\linewidth]{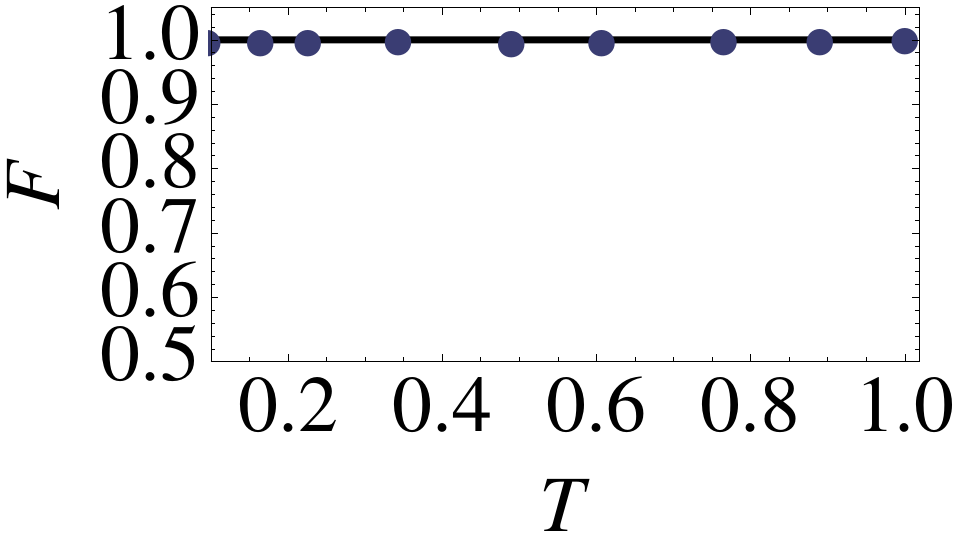}\label{fig:PlotTF}}
\caption{Fidelities versus transmittance in the classical regime for
the two aperture geometries: theoretical predictions (solid lines)
and experimental data (circles).} \label{fig:PlotsTF}
\end{figure}

A comparison between the theoretical curves and the experimental
points in Figs.~\ref{fig:PlotsB}, \ref{fig:PlotrP} and
\ref{fig:PlotsTF} shows a good quantitative agreement. Indeed, as
highlighted in Fig.~\ref{fig:PlotsTF}, even when the position of the
obstruction causes a significant decrease of the transmittance, the
state fidelity remains always above \(\perc{90}\). This demonstrates
that, even in a high-loss regime, the initial information content
encoded in the unperturbed state is preserved in the given OAM
subspace even if a significant spread of the initial OAM spectrum
takes place (see Fig.~\ref{fig:Spectrum}). Any discrepancies may be
explained by an imperfect mode generation and, more appreciably, by
a higher sensitivity to small fluctuations of the fidelity parameter
when \(T\) approaches zero. This result shows that the information
content of OAM qubits exhibits a remarkable resilience to
perturbations such as those here examined.

In particular, the reported high fidelities correspond to the
experimental fact that after the aperture one still has
\(|\kappa_{\psi^\perp}|^2\ll|\kappa_\psi|^2\). This result is tied
to the moderate spread of the OAM spectrum of the perturbed state
even for low values of \(T\) and the progressive shift of the
central spectral state when the initial mode is almost completely
blocked, as shown in Fig.~\ref{fig:Spectrum}.

The present work focuses on qubits encoded in the bidimensional
\(o_2\) subspace. However, it is worth noting that higher-order
subspaces \(o_k\), with LG basis states \(\brc{\ket{+k},\ket{-k}}\),
are likely to offer higher and higher resilience, the higher the OAM
winding number \(k\) (while we may expect a lower resilience in the
$o_1$ subspace). Indeed, as shown in Fig.~\ref{fig:Spectrum} for the
\(\abs{\ell}=2\) case, the increased distance \(\Delta\ell\) between
the two orthogonal basis states causes the spread of the detection
probabilities around a perturbed basis state \(\ket{+k}\) to have a
decreasing overlap with \(\ket{-k}\), as \(k\) increases. Therefore,
while the information preservation probability is strongly affected,
the fidelity is expected to remain high even for very low
transmittance. Of course, these results do not apply to other
subspaces of OAM not involving opposite values of the OAM
eigenvalue, so it remains to be seen whether the same resilience can
be achieved for OAM qudits. Our results are also in qualitative
agreement with former investigations on the OAM spectrum broadening
occurring for LG beams passing through variable angular optical
apertures \cite{Jha:2008,Jack:2009}.

\section{Resilience of hybrid polarization-OAM entanglement} \label{sec:HybridEntanglement}
After the classical regime experiments, we moved to a quantum
regime. We are specifically interested in evaluating the resilience
of the entanglement involving the OAM degree of freedom of a photon,
under the effect of an optical aperture. In particular, we have
considered the case of a planar aperture (function \(B(x_0)\))
inserted in the path of a photon belonging to photon pair that is
initially prepared in a hybrid entangled state of OAM and
polarization. The choice of using hybrid entanglement is mainly
practical, as we start from a polarization-entangled pair of photons
and then the quantum state of one of the two photons is transferred
into OAM by using a polarization-OAM transferrer, as recently
reported in \cite{Nagali:2010}.

\begin{figure}
\includegraphics[width=\linewidth]{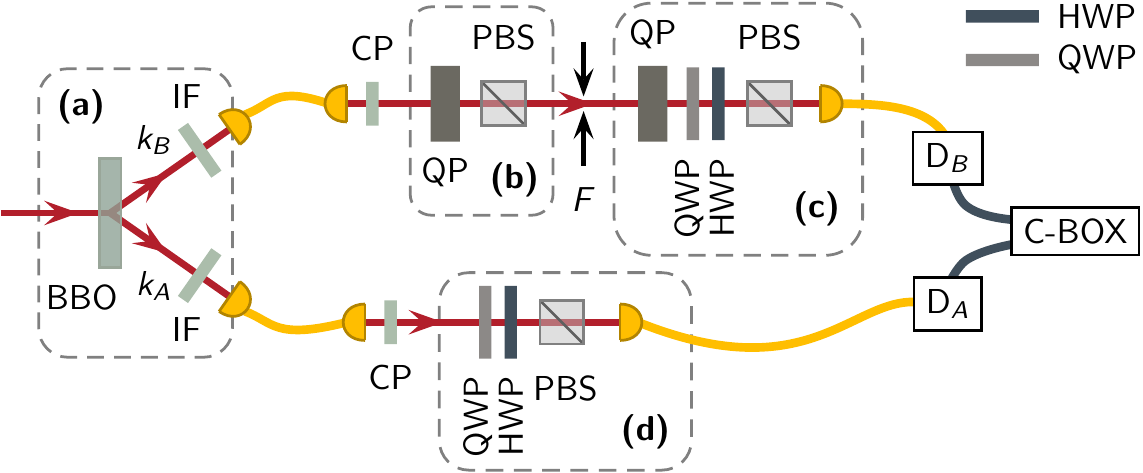}
\caption{Experimental setup for the generation of hybrid
polarization-OAM entangled photon pairs, with the application of the
aperture function \(F\). (a) Generation of polarization-entangled
photons on modes \(k_A\) and \(k_B\). (b) Encoding of the state of
one photon in the OAM subspace \(o_2\) through the \(\pi\to o_2\)
transferrer. (c) OAM analysis. (d) Polarization analysis.}
\label{fig:HybEntSetup}
\end{figure}

Polarization-entangled photon pairs were created by spontaneous
parametric down-conversion; the spatial profile of the twin photons
was filtered through single-mode fibers, and the polarization state
of one of them was coherently transferred to the corresponding OAM
state. The aperture is then used to perturb the photon carrying the
OAM entangled information and, finally, the two-photon quantum state
is analyzed and the fidelity with the initially-prepared state is
computed. The experimental arrangement, shown in
Fig.~\ref{fig:HybEntSetup}, is analogous to that adopted in the
high-quality generation of hybrid polarization-OAM entangled photon
pairs in \cite{Nagali:2010}, and it extends to the case of entangled
photon pairs the setup used in the previous section. A
\(\SI{1.5}{mm}\)-thick \(\beta\)-barium borate crystal (BBO) cut for
type-II phase matching was pumped by the second harmonic of a Ti:Sa
mode-locked laser beam. Via spontaneous parametric fluorescence, the
BBO generated polarization-entangled photon pairs on modes \(k_A\)
and \(k_B\) with wavelength \(\lambda=\SI{795}{nm}\) and pulse
bandwidth \(\Delta\lambda=\SI{4.5}{nm}\), as determined by two
interference filters (IF). The spatial and temporal walk-off was
compensated by inserting a \(\nicefrac{\lambda}{2}\) wave plate and
a \(\SI{0.75}{mm}\)-thick BBO crystal on each output mode \(k_A\)
and \(k_B\) \cite{Kwiat:1999}. The source thus generated photon
pairs in the polarization-encoded singlet entangled state, i.e.
\(\frac{1}{\sqrt{2}}\paren{\ket{H}_A\ket{V}_B-\ket{V}_A\ket{H}_B}\).

The photon in mode \(k_A\) was sent through a standard polarization
analysis setup and then coupled to a single-mode fiber connected to
the single-photon counter modules \(\text{D}_A\). The photon in mode
\(k_B\) was coupled to a single-mode fiber, in order to collapse its
transverse spatial mode into a pure \(\text{TEM}_{00}\)
(\(\ell=0\)). After the fiber output, two wave plates compensated
the polarization rotation introduced by the fiber (CP). To transform
the polarization-entangled pairs into a hybrid entangled state,
photon \(B\) was sent through the quantum transferrer \(\pi\to
o_2\), which converted the polarization quantum states into the
corresponding OAM states. After the transferrer operation the
polarization entangled state is transformed into the hybrid
entangled state
$\frac{1}{\sqrt{2}}(\ket{H}^{A}_{\pi}\ket{+2}^{B}_{o_2}
-\ket{V}^{A}_{\pi}\ket{-2}^{B}_{o_2})$. After the aperture $F$, the
inverse \(o_2\to\pi\) transferrer and a polarization analysis set
was used to analyze the OAM photon state. Ultimately, the photon was
coupled to a single-mode fiber and then detected by \(\text{D}_B\),
connected to the coincidence box (CB), which recorded the
coincidence counts \(\brc{\text{D}_A,\text{D}_B}\).

\begin{figure}
\subfigure[][]{\includegraphics[width=\linewidth]{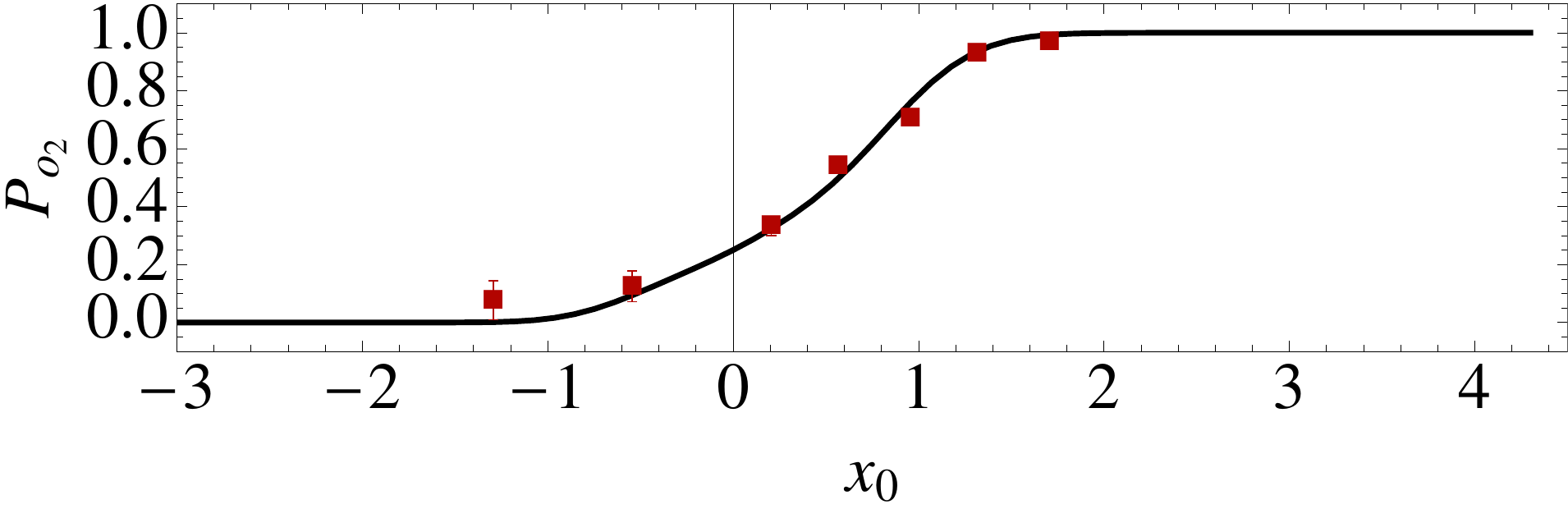}\label{fig:PlotxPEnt}} \\
\subfigure[][]{\includegraphics[width=\linewidth]{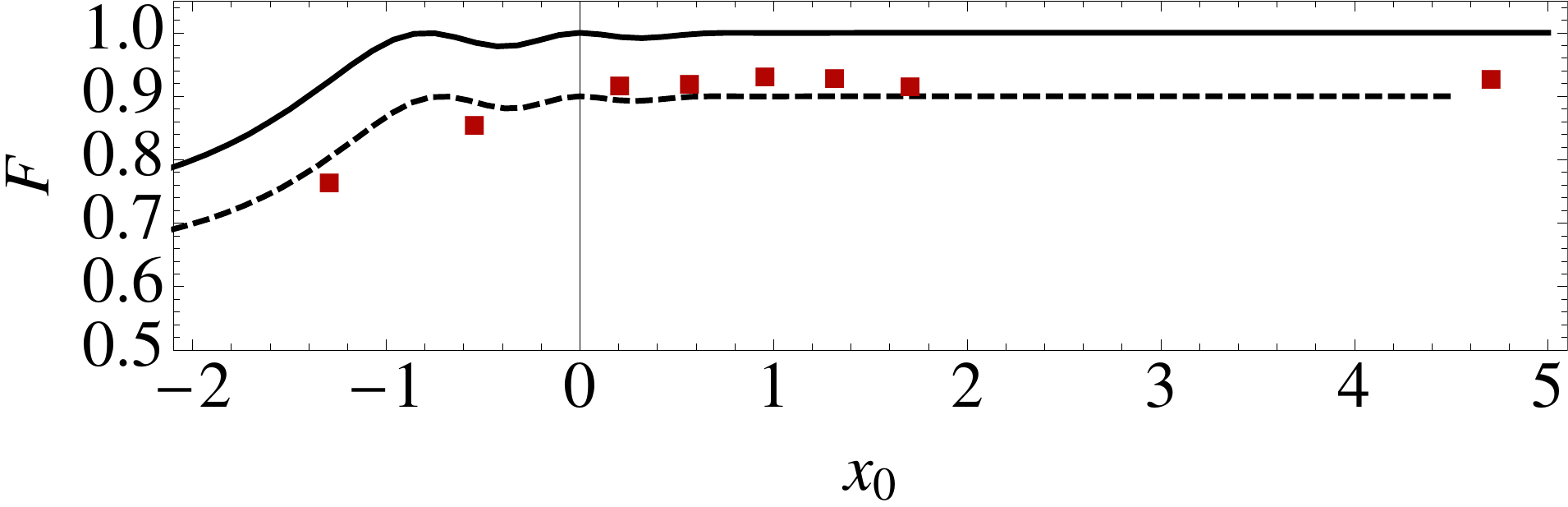}\label{fig:PlotxFEnt}}
\caption{Information preservation probability \(P_{o_2}\)
(\subref{fig:PlotxPEnt}) and fidelity (\subref{fig:PlotxFEnt}) of
the polarization-OAM hybrid-entangled pair of photons as a function
of the planar aperture edge position \(x_0\), given in units of the
beam waist \(w_0\)): theoretical predictions (continuous line) and the one rescaled by the
maximum value of measured fidelity (dashed line) compared to 
experimental data (squares).} \label{fig:PlotsBEnt}
\end{figure}

The experimental results for the probability \(P_{o_2}\) and the
fidelity \(F\) are shown in Fig.~\ref{fig:PlotsBEnt}. We observed
that the state fidelity achieves somewhat lower values than what
predicted (and than what is obtained for the classical experiment
case), particularly when the aperture blocks more than half of the
beam. It is clear from the figure that the fidelity reduction cannot
be entirely ascribed to the aperture, as it remains constant even
when the aperture is moved completely off the beam. The maximum
fidelity of the two photon state is presumably decreased by
imperfections in the source, preparation and measurement stages, to
which is typically more sensitive than the single photon state. To
account for this effect, we plotted the expected theoretical curves
rescaled by the maximum value of measured fidelity. This rescaled
curve exhibits a reasonable agreement with the experimental data
(dashed line in Fig.\ref{fig:PlotsBEnt}b).

\section{Conclusions}
In this paper we tested the robustness of OAM-encoded qubits, which
provide a useful and versatile quantum communication resource. In
the classical case, we first demonstrated that the information
encoded in a bidimensional subspace of OAM can be retrieved
probabilistically in the same subspace even if the state is highly
perturbed in such a way as to block a significant fraction of the
transverse extension of the mode. The experimental results are in
good agreement with the theoretical model.

We then proceeded to demonstrate also in the single-photon regime,
by using polarization-OAM entangled photon pairs, the high
resilience of single-photon bidimensional OAM states. We verified
that hybrid entanglement correlations persist even in high-loss
conditions.

This work was supported by project HYTEQ-FIRB, Finanziamento Ateneo
2009 of Sapienza Universit\`a di Roma, and European project
PHORBITECH of the FET program (Grant No. 255914).

\end{document}